\documentclass[preprint,aps,pra,showpacs,floatfix]{revtex4}
\usepackage{graphicx}
\usepackage{times}
\usepackage {euscript}
\usepackage{amsmath}
\usepackage{amsthm}
\usepackage{epsf}
\usepackage{epsfig}
\usepackage{bm}
\usepackage{bbm}
\renewcommand{\vec}[1]{{\mbox{\boldmath$#1$}}}

\begin{document}
\title{Relativistic calculations 
of the K-K charge transfer and K-vacancy production probabilities in low-energy ion-atom collisions}
%
%
\author{ I.~I.~Tupitsyn,$^1$  Y.~S.~Kozhedub,$^1$ V.~M.~Shabaev,$^1$
A.~I.~Bondarev,$^1$ G.~B.~Deyneka,$^2$
 I.~A.~Maltsev,$^1$
S.~Hagmann,$^{3}$ G.~Plunien$^4$, and Th.~St\"ohlker$^{3,5,6}$}
\affiliation{
$^1$
Department of Physics, St. Petersburg State University, Oulianovskaya 1,
Petrodvorets, St. Petersburg 198504, Russia \\
$^2$
St. Petersburg State University of Information Technologies,
Mechanics and Optics, Kronverk av. 49, 197101 St. Petersburg, Russia \\
$^3$
Gesellschaft f\"ur Schwerionenforschung,
Planckstrasse 1, D-64291 Darmstadt, Germany \\
$^4$
Institut f\"ur Theoretische Physik, Technische Universit\"at Dresden,
Mommsenstra{\ss}e 13, D-01062 Dresden, Germany \\
$^5$
Physikalisches Institut, Universit\"at Heidelberg, Philosophenweg 12, D-69120
Heidelberg, Germany \\
$^6$
Helmholtz-Institut Jena, D-07743 Jena, Germany \\
}
%
\begin{abstract}
The previously developed technique 
for evaluation of charge-transfer and electron-excitation processes in low-energy heavy-ion 
collisions [I.I. Tupitsyn et al., Phys. Rev. A 82, 042701(2010)] is extended  
to collisions of ions with neutral atoms. The method 
employs  the active electron approximation, in which only
the active electron participates
in the charge transfer and excitation processes while the passive
electrons provide the screening DFT potential. The time-dependent Dirac wave
function of the active electron is represented as a linear combination
of atomic-like Dirac-Fock-Sturm orbitals, localized at the ions (atoms). 
The screening
DFT potential is calculated using the overlapping densities of each
ions (atoms), derived from the atomic orbitals of the passive electrons.
The atomic orbitals are generated by solving numerically the one-center
Dirac-Fock and Dirac-Fock-Sturm equations by means of a finite-difference
approach with the potential taken as the sum of the exact reference ion (atom)
Dirac-Fock potential and of the  Coulomb potential from the other ion within the
monopole approximation. The method developed is used to calculate
the K-K charge transfer and K-vacancy production probabilties for the
Ne$(1s^22s^22p^6)$~--~F$^{8+}(1s)$ collisions at the F$^{8+}(1s)$
projectile energies
$130$ keV/u and $230$ keV/u. The obtained results are compared with
experimental data and other theoretical calculations. The K-K charge transfer
and K-vacancy production  probabilities are also calculated for the
Xe~--~Xe$^{53+}(1s)$  collision.
\end{abstract}

\maketitle
%
%
\section{Introduction}

Collisions of highly charged ions provide a unique tool
for tests of relativistic and quantum
electrodynamics (QED) effects
in the scattering theory~\cite{Eichler_95,Shabaev_02,Eichler_07}.
Investigations of such
processes can also give an access
to  QED in supercritical
 fields, provided the total charge of the colliding nuclei is
larger than the critical one, $Z_c=173$ (see, e.g., 
Ref.~\cite{Greiner_85} and references therein).
One of the most attractive way for indirect observation of the
supercritical field created in the collision
 is to investigate the 
 dynamics of inner-shell electrons that can be 
 rather sensitive to the collision parameters. 
The most favorable conditions  for studying the electron dynamics 
in the supercritical field regime
 correspond to the projectile energy of about
the Coulomb barrier \cite{mul_76}. In case of U$-$U collision this requires
the energy of about $6$~MeV/u that means a low-energy collision.
One of the key processes in such collisions is the charge transfer of 
electrons (see, e.g., Ref.~\cite{Bransden_92} and references therein). 
A systematic approach to relativistic calculations of the charge-transfer
and electron-excitation 
probabilities in low-energy heavy-ion collisions was developed
in our previous paper~\cite{Tupitsyn_10}, where the consideration was
restricted to collisions of H-like ions with bare nuclei.
Since the experimental study of such collisions for high-$Z$ systems
is presently 
rather problematic, an extention of the method to collisions
of highly charged ions with neutral atoms, that can be 
studied in experiments with the current GSI and future FAIR 
facilities~\cite{Hagmann_10, Hagmann_11}, is needed.
In this paper
we present the desired extention and perform calculations 
for  low-energy ion-atom collisions.
To examine the approach we calculate the K-K charge transfer and K-vacancy 
production probabilities for  low-energy collision of H-like F (F$^{8+}$) 
and neutral Ne, the process which has been investigated
both  experimentally~\cite{Hagmann_87}
and theoretically~\cite{Fritsch_85,Toepfer_87,Thies_89}.
The calculations are performed
 at the F$^{8+}$ projectile energies $130$~keV/u and $230$~keV/u.
We also evaluate the probabilities of
 the K-K charge transfer and K-vacancy production 
in the Xe~--~Xe$^{53+}(1s)$ collision at 
the projectile
energy of $3.6$~MeV/u. The latter processes are planned to be
studied in the nearest future in
experiments at GSI ~\cite{Hagmann_10, Hagmann_11}. 

The paper is organized as follows.
In the section~\ref{subsec:Kohn-Sham} we describe the time-dependent
one-electron  equation in so-called active electron approximation~\cite{Lin_81}
and the method for constructing the local Kohn-Sham
potential induced by the passive electrons. The wave function of
the active electron is  expanded in terms of the Dirac-Fock and Dirac-Fock-Sturm
basis functions, which are central-field 4-component Dirac bispinors
centered at the ions.
The two-center relativistic Kohn-Sham equation in the finite basis set
is briefly discussed in the section~\ref{subsec:two-center}.
The basis functions are obtained by solving numerically the
atomic Dirac-Fock and Dirac-Fock-Sturm equations in an external field with
a special choice of the weight function, as it was proposed in
Refs.~\cite{Tupitsyn_03, Tupitsyn_05}. The external potential is spherically
symmetric Coulomb-Hartree potential of the other ion (atom) taken in the
monopole approximation. The basis set constructed in this way
and the related calculation procedures
are described in section~\ref{subsec:basis}.
Basic formulas for the K-vacancy production probability 
are given in section~\ref{subsec:prob}.
In section~\ref{subsec:results} we present the results of the
relativistic calculations of the Ne K-shell-vacancy production
and K-K charge transfer probabilities in
the Ne~--~F$^{8+}$ collisions
as a function of the impact
parameter $b$ at the projectile energies $130$~keV/u and $230$~keV/u.
In this section we also present the results of the neutral 
Xe K-shell-vacancy production and K-K charge transfer probabilities
in the Xe~--~Xe$^{53+}(1s)$ collision.
%
\section{Theory}
\label{sec:theory}
%
\subsection{ Dirac-Kohn-Sham equation in the active electron approximation}
\label{subsec:Kohn-Sham}
%
In this paper we use so-called active electron approximation~\cite{Lin_81}
to describe the ion-atom collision. In this
approximation, we consider only the active electron which participates in
the charge transfer and excitation processes, while the other passive
electrons provide a screening potential. In our
calculations the screening potential is  defined by the density functional
theory (DFT) in the local density approximation (LDA).  In this approach
the time-dependent wavefunction $\psi(\vec{r},t)$ of the active electron
$\psi(\vec{r},t)$ is the solution of the relativistic time-dependent
Kohn-Sham equation. In atomic units ($\hbar=m=e=1$), this equation is
given by
\begin{equation}
i \frac{\partial \psi(\vec{r},t)}{\partial t} = h_{\rm D} \,
\psi(\vec{r},t) \,.
\end{equation}
Here $h_{\rm D}$ is the two-center Dirac-Kohn-Sham Hamiltonian defined by
\begin{equation}
\hat h_{\rm D} =c (\vec{\alpha} \cdot \vec{p}) +
\beta \, c^2 + V_{AB}(\vec{r}) \,, \qquad
V_{AB}(\vec{r})= V_{H}[\rho] + V_{xc}[\rho]\,,
\end{equation}
where $c$ is the speed of light and $\vec{\alpha}$, $\beta$ are the
Dirac matrices. $V_{H}[\rho]$ and $V_{xc}[\rho]$ are the Hartree and
exchange-correlation potentials, respectively. Both of them are
the functionals of the electron density $\rho(\vec{r})$.
The Hartree potential $V_{H}[\rho]$ includes the electron-nucleus 
interaction and the electron-electron Coulomb repulsion $V_C[\rho]$:
\begin{equation}
V_{H}(\vec{r}) = V_{\rm nucl}^{A}(\vec{r}_A) + V_{\rm nucl}^{B}(\vec{r}_B) +
V_C[\rho]
\,, \qquad \vec{r}_A=\vec{r}- \vec{R}_A\,, \qquad \vec{r}_B=\vec{r}- \vec{R}_B,
\end{equation}
where
\begin{equation}
V_{\rm nucl}(\vec{r}) =
\int d^3 \vec{r}^{\prime} \, 
\frac{\rho_{\rm nucl}(\vec{r}^{\prime})}{|\vec{r}-\vec{r}^{\prime}|} \,,
\qquad V_C[\rho] = \int \, d^3\vec{r^{\prime}} \, 
\frac{\rho(\vec{r}^{\prime})}{|\vec{r}-\vec{r}^{\prime}|} \,,
\end{equation}
$\rho_{\rm nucl}(\vec{r})$ and $\rho(\vec{r})$
are the nuclear and electron densities, respectively. The exchange-correlation
potential $V_{xc}[\rho]$ was taken in the Perdew-Zunger
parametrization~\cite{Perdew_81} including the self-interaction correction (SIC).

The electron density $\rho(\vec{r})$, obtained with the many-electron
wave function which is represented by a Slater determinant, is invariant 
with respect to the rotations
in the occupied orbitals space. For this reason, to obtain the
electron density we can use atomic-like orbitals, localized on the both
centers
($A$ and $B$). In this case the electron density, constructed from the
orthogonal localized orbitals, can be represented as a sum of densities
$\rho_A(\vec{r})$ and $\rho_B(\vec{r})$ which are localized on 
the centers $A$ and $B$.
This is not the case, however, if the orbitals localized on the different centers overlap and
are non-orthogonal. The electron density, derived from non-orthogonal
orbitals, is given by
\begin{equation}
\rho(\vec{r}) = \sum_{i,j} \psi_i^{(p)^{\ast}}(\vec{r}) \,
\left(S^{-1} \right)_{ij} \, \psi_j^{(p)}(\vec{r}) \,,
\end{equation}
where $\psi_i^{(p)}(\vec{r})$ are the atomic-like wavefunctions of the
passive electrons and matrix $S$ is the overlapping matrix. Note that the electron
density is normalized on the number of the passive electrons,
\begin{equation}
\int d^3 \vec{r} \, \rho(\vec{r}) = N-1 \,,
\end{equation}
where $N$ is the total number of electrons.
The electron density $\rho(\vec{r})$ can be divided into three parts,
\begin{equation}
\rho(\vec{r}) = \rho_A(\vec{r}) +  \rho_B(\vec{r}) + 
\rho^{\rm (ovlp)}_{AB}(\vec{r}) \,,
\end{equation}
if we split the summation over indices $i,j$ into the sum over $i,j \in A$,
the sum over $i,j \in B$ and the remaining overlapping part.
We can also split the overlapping density into two parts dividing the space
into two regions $(A)$ and $(B)$,
\begin{equation}
\rho^{\rm (ovlp)}_{AB}(\vec{r}) =
\rho^{\rm (ovlp)}_{A}(\vec{r})+\rho^{\rm (ovlp)}_{B}(\vec{r}).
\end{equation}
This can be done by the plane passing through
the middle of the internuclear distance (see
Ref.~\cite{Tupitsyn_10} for details).

For simplicity, let us consider
spherically-average values of
the electron densities in each region,
\begin{equation}
\overline \rho_A(r_A) = \int d\Omega_A \,
\left[ \rho_A(\vec{r})+\rho^{\rm (ovlp)}_{A}(\vec{r}) \right ]\,, \qquad
\overline \rho_B(r_B) = \int d\Omega_B \,
\left [\rho_B(\vec{r})+\rho^{\rm (ovlp)}_{B}(\vec{r}) \right] \,.
\end{equation}
This procedure does not change the normalization of the total electron
density,
\begin{equation}
\int d^3{\vec{r}} \, \overline \rho(\vec{r}) = 
\int d^3{\vec{r}} \, \left [\overline \rho_A(r_A)+
\overline \rho_B(r_B) \right ] = N-1 \,.
\end{equation}
As a result, the potential $V_{AB}(\vec{r})$ can be approximated
by  the sum of the spherically symmetric potentials of the two different centers,
\begin{equation}
V_{AB}(\vec{r}) \simeq V_A[\overline \rho_A](r_A) + 
V_B[\overline \rho_B](r_B) \,.
\end{equation}

The overlapping densities must be taken 
into account especially for the short internuclear distances. Otherwise, the Pauli
principle is violated and, as a result, the
number of electrons on the $1s$ shell of the united system 
can exceed $2$.

It should also be noted that the time-dependent wave function $\psi(\vec{r},t)$ 
of the active
electron is orthogonalized to the wave functions $\psi^{(p)}_i(\vec{r})$
of the passive  electrons.
This means that the transitions of the active electron to the states
occupied by
the passive electrons are forbidden in accordance with the Pauli principle.

%
\subsection{ Two-center Dirac-Kohn-Sham equation}
\label{subsec:two-center}
The two-center expansion of the time-dependent wave function $\psi(\vec{r},t)$
can be written in the form
\begin{equation}
\psi(\vec{r},t)  =  \displaystyle \sum_{\alpha=A,B} \, \sum_{a}
C_{\alpha a}(t) \, \varphi_{\alpha,a} (\vec{r}-\vec{R}_\alpha(t)) \,,
\label{expan1}
\end{equation}
where index $\alpha=A,B$ labels the centers, index $a$
enumerates basis functions at the given center, and
$\varphi_{\alpha,a} (\vec{r}-\vec{R}_\alpha)$ is the central-field
bispinor centered at the point $\alpha$.
In what follows,  the shorthand notations
$|j\rangle \equiv |\varphi_j\rangle \equiv |\varphi_{\alpha , a}\rangle$
for states $j \equiv \alpha , a$ are used.
The expansion coefficients $C_{a \alpha}(t)$ of the time-dependent
wave function $\psi(\vec{r},t)$ can be obtained by  solving
the linear system of first-order differential equations
\begin{equation}
i \sum_{k}S_{jk} \frac{dC_{k}(t)}{d t}  =  \sum_{k} \,
( H_{jk} - T_{jk}) \, C_{k}(t)  \,,
\end{equation}
where indices $j$ and $k$ enumerate the basis functions of both centers, and
the matrix elements of $H$ and $S$ are 
\begin{equation}
H_{jk} \,=\, \langle j \mid \hat h_{\rm D} \mid k \rangle \,, \qquad
S_{j k} \,=\, \langle j \mid k \rangle \,.
\label{matr1}
\end{equation}
The matrix elements of $T$ are given by
\begin{equation}
T_{jk} \,=\, i \langle j \mid \frac{\partial}{\partial t} \mid k \rangle =
T^{\ast}_{kj} + i \frac{\partial}{\partial t} \, S_{jk} \,.
\label{matr2}
\end{equation}
Obviously the matrix $T$ is non-Hermitian, if the overlapping matrix $S$
depends on time.

The functions $\varphi_{\alpha , a}$ depend on time due to two reasons. 
First, the basis functions centered at the target and projectile nuclei
move together with the nuclei. Second, the basis functions depend
parametrically on the distance between the nuclei, since their radial
parts are obtained from the radial equations, where for each center
the potential of the other ion (atom) is included in the so-called monopole
approximation (see section~\ref{subsec:basis}).

Calculations of the matrix elements $H_{jk}$,
$S_{jk}$, and $T_{jk}$ were considered in detail in Ref.~\cite{Tupitsyn_10}.
%
\subsection{Basis functions}
\label{subsec:basis}
In our approach the basis set contains Dirac-Fock and Dirac-Fock-Sturm
orbitals. The Dirac-Fock-Sturm orbitals can be considered as pseudo-states,
which should be included in the basis to take into account the
contribution of the positive- and negative-energy Dirac continuum.
Both types of basis functions $\varphi_{\alpha a}$ are central field
Dirac bispinors centered at the position 
$\vec{R}_{\alpha}$ ($\alpha=A,B$) of the corresponding ion,
\begin{equation}
\varphi_{n\kappa m}(\vec{r}) =
\left  ( \begin{array}{l} \displaystyle
\,\, \frac{~P_{n \kappa}(r)}{r} \,  \chi_{\kappa m}(\Omega)
\\[4mm] \displaystyle
i \, \frac{Q_{n \kappa}(r)}{r} \, \chi_{-\kappa m}(\Omega)
\end{array} \right ) \,,
\end{equation}
where $P_{n \kappa}(r)$ and $Q_{n \kappa}(r)$ are the large and small radial
components, respectively, and $\kappa=(-1)^{l+j+1/2}(j+1/2)$ is the
relativistic angular quantum number.
The large and small radial components are obtained by solving
numerically the Dirac-Fock and Dirac-Fock-Sturm equations in the central
field approximation. The radial Dirac-Fock equation is
\begin{equation}
\left(h^{\rm DF}_{\alpha} + V_{\rm ext}(r) \right) \, F_{\alpha n\kappa}(r)
= 
\varepsilon_{\alpha n\kappa}  F_{\alpha n\kappa}(r) \,, \qquad
F_{\alpha n\kappa}(r) = \left(
\begin{array}{c}
P_{\alpha n\kappa}(r) \\ Q_{\alpha n\kappa}(r)
\end{array}
\right ),
\label{dirac1}
\end{equation}
where $ h^{\rm DF}_{\alpha}$ is the radial Dirac-Fock Hamiltonian of ion $\alpha$
($\alpha=A,B$),  $F_{\alpha n\kappa}(r)$ is the two-component radial wave function,
and $V_{\rm ext}(r)$ is a local external potential. The explicit form
of the radial Dirac-Fock equation and the description of the corresponding
computer code
are presented in Ref.~\cite{Bratsev_77}. 
The radial components of the Dirac-Fock-Sturm orbitals
$\overline{\varphi}_{n\kappa m}$, which we denote by
$\overline F_{n\kappa}(r)$, are the solutions of the generalized
Dirac-Fock-Sturm eigenvalue problem,
\begin{equation}
\left(h^{\rm DF}_{\alpha} + V_{\rm ext}(r) -
\varepsilon_{\alpha n_0 \kappa} \right ) \, \overline F_{\alpha n\kappa}(r) = 
\lambda_{\alpha n \kappa} \, W_{\alpha \kappa}(r) \,
\overline F_{\alpha n \kappa}(r) \,.
\label{sturm1}
\end{equation}
Here $\lambda_{\alpha n \kappa}$ can be considered as the eigenvalue of the
Dirac-Fock-Sturm operator and $W_{\alpha \kappa}(r)$ is a constant sign weight
function. The energy $\varepsilon_{\alpha n_0 \kappa}$ is fixed in the
Dirac-Fock-Sturm equation. If the weight function $W(r) \to 0$ at $r \to \infty$,
all Sturmian functions have the same asymptotic behavior at $r \to \infty$.
It is clear that for $\lambda_{\alpha n \kappa}=0$ the Sturmian function
$\overline{\varphi}_{n\kappa m}$  coincides with the reference Dirac-Fock orbital
$\varphi_{\alpha n_0 \kappa}$. In our calculations we use the following weight
function
\begin{equation}
W_{\kappa}(r)  \,=\, - \, \frac{1 \,-\, 
\exp(-(\alpha_{\kappa} \, r)^2)}{(\alpha_{\kappa} \, r)^2}\,.
\label{sturm2}
\end{equation}
In contrast to $1/r$, this weight function is regular at the origin.
It is well-known that the Sturmian operator is Hermitian. It does
not contain continuum spectra, in contrast to the Dirac operator.
Therefore, the set of the Sturmian eigenfunctions forms a discrete
and complete basis set of one-electron wave functions.

The external central-field potential $V_{\rm ext}(r)$ in equations
(\ref{dirac1}) and (\ref{sturm1}) is arbitrary, and, therefore, it can be chosen 
to provide most appropriate Dirac-Fock and Dirac-Fock-Sturm basis orbitals. 
At small internuclear distances the wave function of the atomic electron
experiences also the strong Coulomb field of the other ion (atom).
To the leading order this effect can be taken into account by 
including the Coulomb-Hartree potential of the second ion (atom)
as the external potential $V_{\rm ext}(r)$ within the so-called monopole
approximation. For instance, the external central-field potential
$V^{A}_{\rm ext}(r)$  is given by
\begin{equation}
V^{A}_{\rm ext}(r) = V^{B}_{\rm mon}(r) = 
\frac{1}{4 \pi} \,\int d\Omega_A \,\, V^{B}_{H}(\vec{r}-\vec{R}_{AB}) \,. 
\end{equation}
where $V^{B}_{\rm mon}(r)$ is the spherically-symmetric part of the
reexpansion of the Coulomb-Hartree potential $V^{B}_{H}(\vec{r}-\vec{R_{AB}})$ of the
ion $B$ with respect to the center $A$ and 
$\vec{R}_{AB}$ is the internuclear distance vector. 

Calculations of the two-center integrals with the basis
functions employed were described in detail
in our previous paper~\cite{Tupitsyn_10}.
The time-dependent Dirac-Kohn-Sham equation for the active electron is
solved using the two-center basis set expansion.
The expansion coefficients are determined employing the direct evolution
(exponential) operator method~\cite{Tupitsyn_10}, which is more stable
compared to the others, such as, e.g., the Crank-Nicholsen propagation
scheme~\cite{Crank_47} and the split-operator method~\cite{Fett_82}.
To obtain the matrix representation of the exponential operator
in the finite basis set one has to diagonalize the generalized complex
Hamiltonian matrix at each time step. Since our basis set is not too large,
the diagonalization procedure is not too time consuming.

%
\subsection{Charge-transfer and vacancy-production probabilities}
\label{subsec:prob}
%
The amplitudes of the charge transfer and excitations to different
bound states of the projectile and target ions are calculated
by projecting the time-dependent wave function
of the active electron onto the atomic Dirac-Fock orbitals of the projectile
and target. The corresponding calculations for collisions 
of H-like ions with bare nuclei
were described in detail in
our work~\cite{Tupitsyn_10}. That is
why here we restrict our consideration only to the new features 
of the calculation procedure that occur for atom-ion 
collisions  within the one active electron approximation.

Consider the collision of a neutral atom $A$ (target)  
with a hydrogenlike ion $B$ (projectile). We assume that before
the collision  the active electron  occupies  the $1s$ state of the target
with spin up (in the relativistic case, with the total angular
moment projection  $\mu=1/2$) and the passive electron of the hydrogenlike ion
occupies the $1s$ state with spin down. In what follows, we are interested in
two processes: the K-K charge transfer and the K-shell vacancy production.

Let $P_A(1s)$ is the probability to find the active electron in the $1s$
state of the target after the collision, $P_B(1s)$ denotes the probability
to find one active electron in the $1s$ state of the projectile or,
in other words, the probability $P_{\rm K-K}$ of the K-K shell charge
transfer of one electron. To obtain the probabilty $P_{\rm vac}$ of the K-shell
vacancy production, we introduce the probabilities $P(E_1)$ and $P(E_2)$ of
the events $E_1$ and $E_2$, when a hole is created in the $1s$ state
of the target with spin up and down, respectively. The probabilities
of these events are defined by
\begin{align}
\left \{
\begin{array}{lll} \displaystyle
P(E_1) &=& \displaystyle 1-P_A(1s) \\[3mm]
P(E_2) &=& \displaystyle 1-P_A(1s)-P_B(1s)
\end{array} 
\right ..
\end{align}
Assuming the events $E_1$ and $E_2$ are independent, the
probability of production at least one hole in the $1s$ state of the target
is given by
\begin{align} \label{pvac}
P_{\rm vac}=P(E_1)+P(E_2)-P(E_1)P(E_2)=
1-P_A(1s)(P_A(1s)+P_B(1s)) \,.
\end{align}
We note that, since the sum of the probabilities $P_A(1s)$ and $P_B(1s)$ is less
than 1, the vacancy production probability satisfies the condition
$0 \le P_{\rm vac} \le 1$. It should also be noted that the K-shell vacancy production
defined by Eq. (\ref{pvac}) includes the production of two holes in the target K shell.

\subsection{Results of the calculations and discussion}
\label{subsec:results}
%
To test the approach we have studied the 
Ne~--~F$^{8+}(1s)$ collision for low energies where
experimental and nonrelativistic theoretical results are
available~\cite{Fritsch_85, Hagmann_87, Toepfer_87, Thies_89}.
In this case the nuclear charge numbers are rather small
and, therefore, relativistic effects are negligible. We stress, however,
that our approach
can be directly applied to heavier systems where the relativistic effects
become stronger or even dominant.

\begin{figure}[hbt]
\centering
\vspace{-0.2cm}
\includegraphics[width=10.5cm,clip]{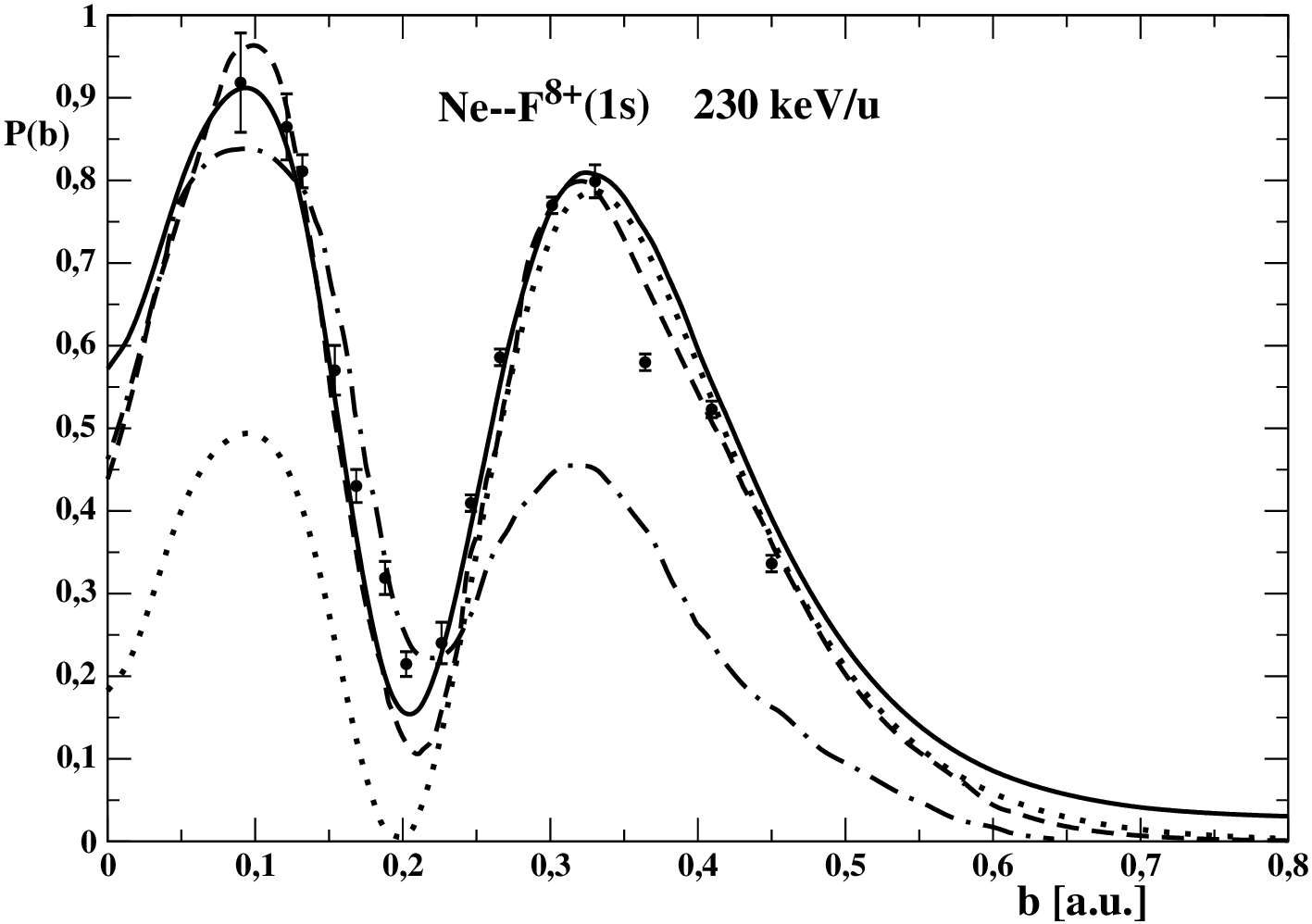}
\caption{\small The results of the present calculations
 for the probabilities $P(b)$ of the Ne
K-shell-vacancy production (solid line) and of the K-K-shell charge transfer 
(dotted line) as  functions of the
impact parameter $b$ for the Ne-F$^{8+}(1s)$ collision at the projectile
energy of $230$ keV/u.  The circles indicate experimental results by Hagmann 
{\it et al.}~\cite{Hagmann_87}.
The dashed and dash-dotted lines present  
theoretical results by  Fritsch and Lin~\cite{Fritsch_85} and by
Thies {\it et al.}~\cite{Thies_89}, respectively.}
\label{Fig:NeF_230}
\end{figure}

In Fig.~\ref{Fig:NeF_230} we present the results of our calculations for
the probabilities $P(b)$ of the Ne
K-shell-vacancy production (solid line) and of the K-K-shell charge transfer 
(dotted line) as  functions of the
impact parameter $b$ for the Ne-F$^{8+}(1s)$ collision at the projectile
energy of $230$ keV/u. The results
for the Ne K-vacancy production are compared with experimental
values (circles)~\cite{Hagmann_87} and with theoretical results obtained
by Fritsch and Lin (dashed line)~\cite{Fritsch_85} and by
Thies {\it et al.} (dash-dotted line)~\cite{Thies_89}.
It can be seen that our results are in perfect agreement with
the experimental ones. 
%

\begin{figure}[hbt]
\centering
\vspace{-0.2cm}
\includegraphics[width=10.5cm,clip]{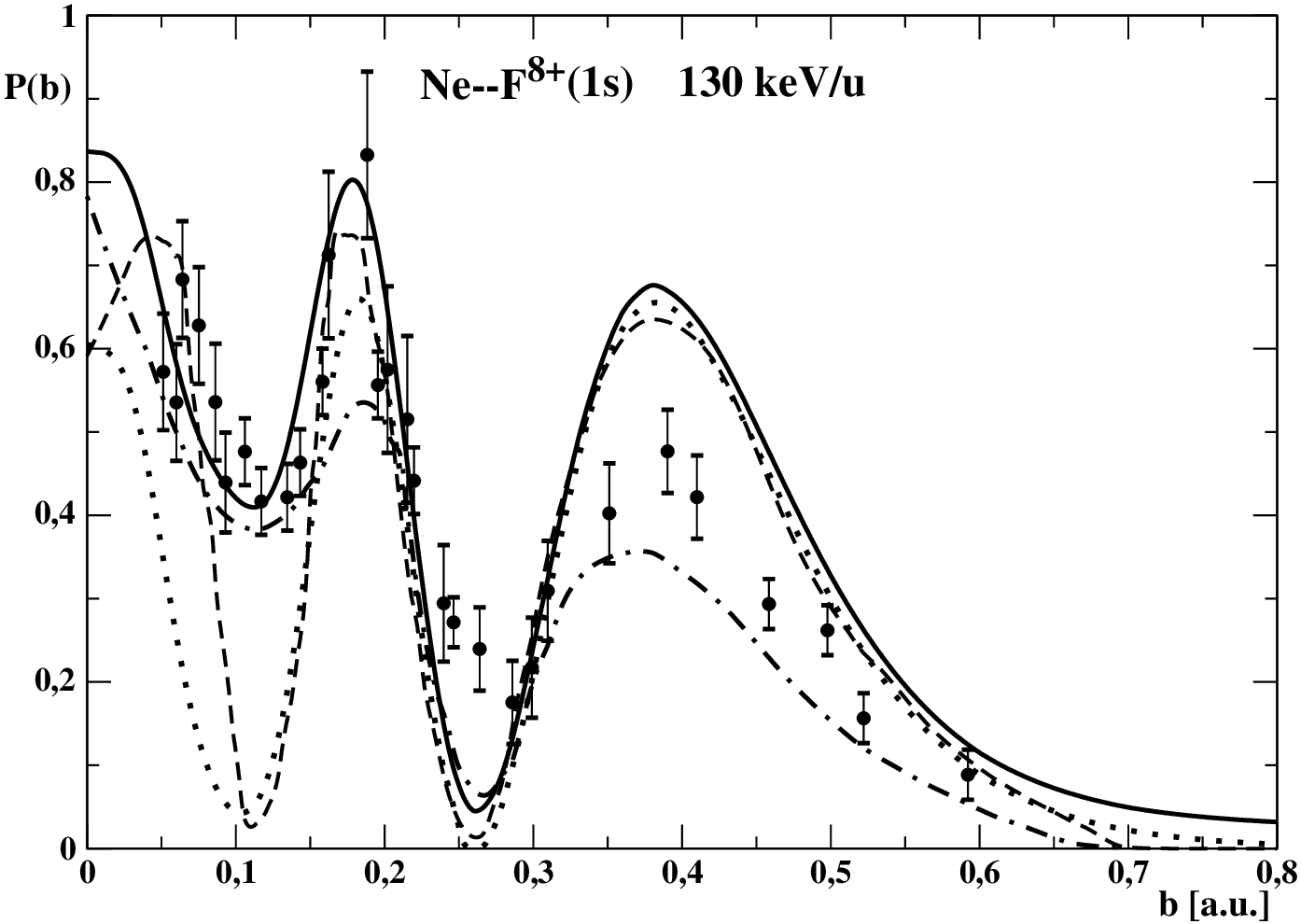}
\caption{\small
 The results of the present calculations for the probabilities $P(b)$ of the Ne
K-shell-vacancy production (solid line) and of the K-K-shell charge transfer 
(dotted line) as  functions of the
impact parameter $b$ for the Ne-F$^{8+}(1s)$ collision at the projectile
energy of $130$ keV/u.  The circles indicate experimental results by Hagmann 
{\it et al.}~\cite{Hagmann_87}.
The dashed and dash-dotted lines present  
theoretical results by  Lin {\it et al.}~(taken from Ref. \cite{Hagmann_87}) and by
Thies {\it et al.}~\cite{Thies_89}, respectively.}
\label{Fig:NeF_130}
\end{figure}

The related results for the Ne-F$^{8+}(1s)$ collision at the projectile 
energy of  $130$ keV/u are presented in  Fig.~\ref{Fig:NeF_130}, where
the same notations as in Fig.~\ref{Fig:NeF_230} are used.
We note that our theoretical results are in a good agreement
with the experimental ones at small impact parameters. 
However, in contrast to Fig. 1, the agreement is not so good 
for medium and large impact parameters, although 
the theoretical predictions for the maximum and minimum
positions agree rather well with the experimental ones.

As one can see from  Figs.~\ref{Fig:NeF_230} and \ref{Fig:NeF_130},
for both energies at large impact parameters
the K-vacancy production is mainly determined by the K-K-shell charge transfer,
which is indicated by the dotted line. The difference between the K-vacancy
production and the K-K-shell transfer probabilities at small impact
parameters is due to the contribution from the charge-transfer excitation
into the $2s$, $2p$, and higher vacant states of the projectile. 
This is in accordance with
the experimental results of Ref. \cite{Hagmann_82},
where the K-vacancy production in the Ne~--~F$^{6+}((1s)^22s)$ collision
was studied. The calculation of the latter process is currently under way
and will be published elsewhere.

\begin{figure}[hbt]
\centering
\vspace{-0.2cm}
\includegraphics[width=10.5cm,clip]{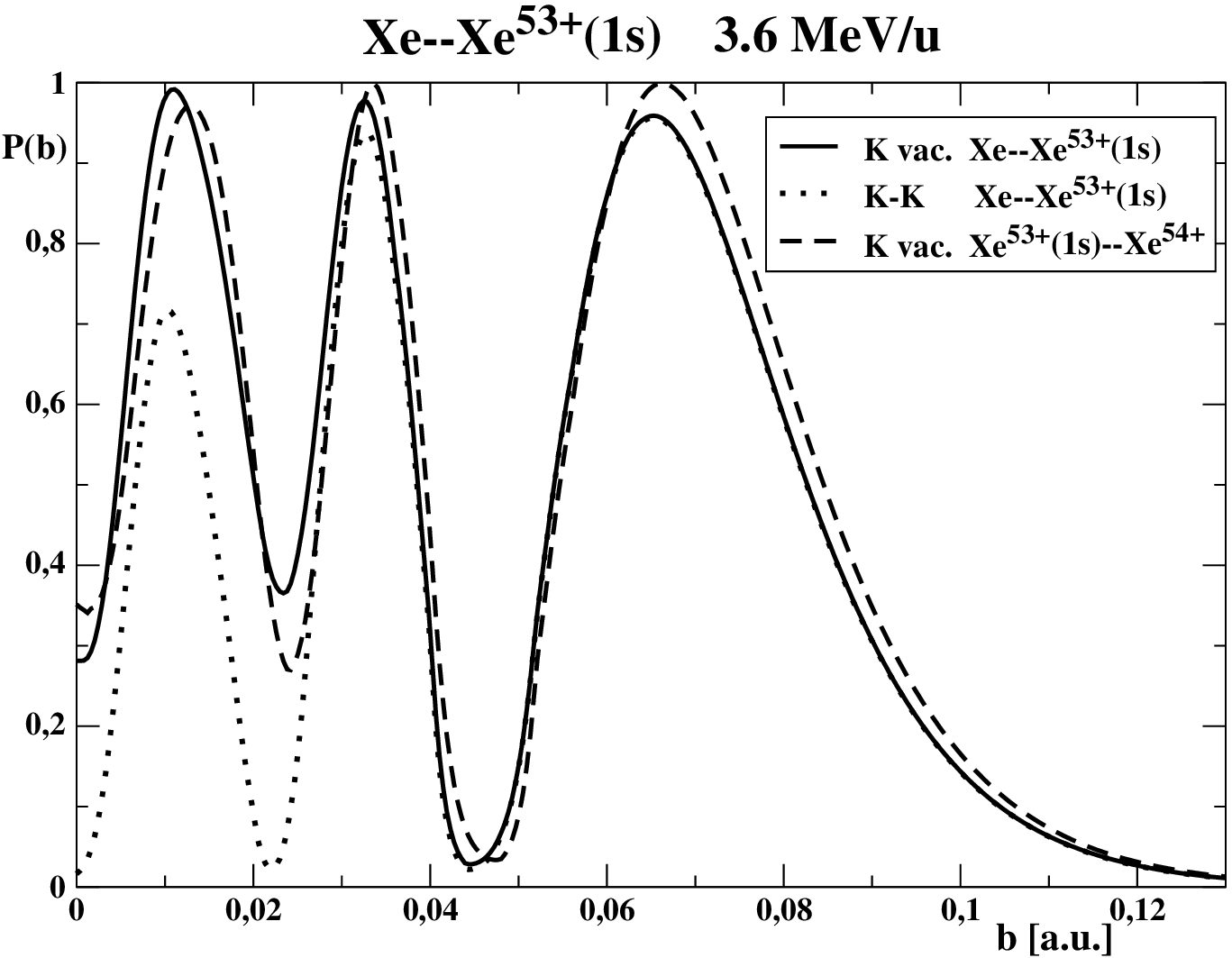}
\caption{\small
The probabilities $P(b)$ of the Xe
K-shell-vacancy production (solid line)
and of the K-K-shell charge transfer (dotted line)
in the  Xe-Xe$^{53+}(1s)$ collision
as functions of the
impact parameter $b$. The dashed line indicates
the K-shell-vacancy production
for the Xe$^{53+}(1s)$-Xe$^{54+}$ collision.}
\label{Fig:XeXe_3.6}
\end{figure}

In this work we also performed the related calculations for
the Xe~--~Xe$^{53+}(1s)$ collision  at the projectile  energy of $3.6$ MeV/u. 
The experimental study of this process is planned at  GSI (Darmstadt)
\cite{Hagmann_10,Hagmann_11}.
The probabilities $P(b)$ of the Xe
K-shell-vacancy production and of the K-K-shell charge transfer 
as functions of the
impact parameter $b$ are plotted in Fig.~\ref{Fig:XeXe_3.6}. The solid and
dotted lines represent the vacancy production and the charge transfer, respectively.
For comparison, in the same figure we display the K-shell-vacancy production
for the Xe$^{53+}(1s)$~--~Xe$^{54+}$ collision that is indicated by 
the dashed line.
We note again that  at large
impact parameters the K-vacancy production is
almost completely determined by
the K-K-shell charge transfer. It can be also seen that
in the case under consideration the screening effect is rather small.

\begin{figure}[hbt]
\centering
\vspace{-0.2cm}
\includegraphics[width=10.5cm,clip]{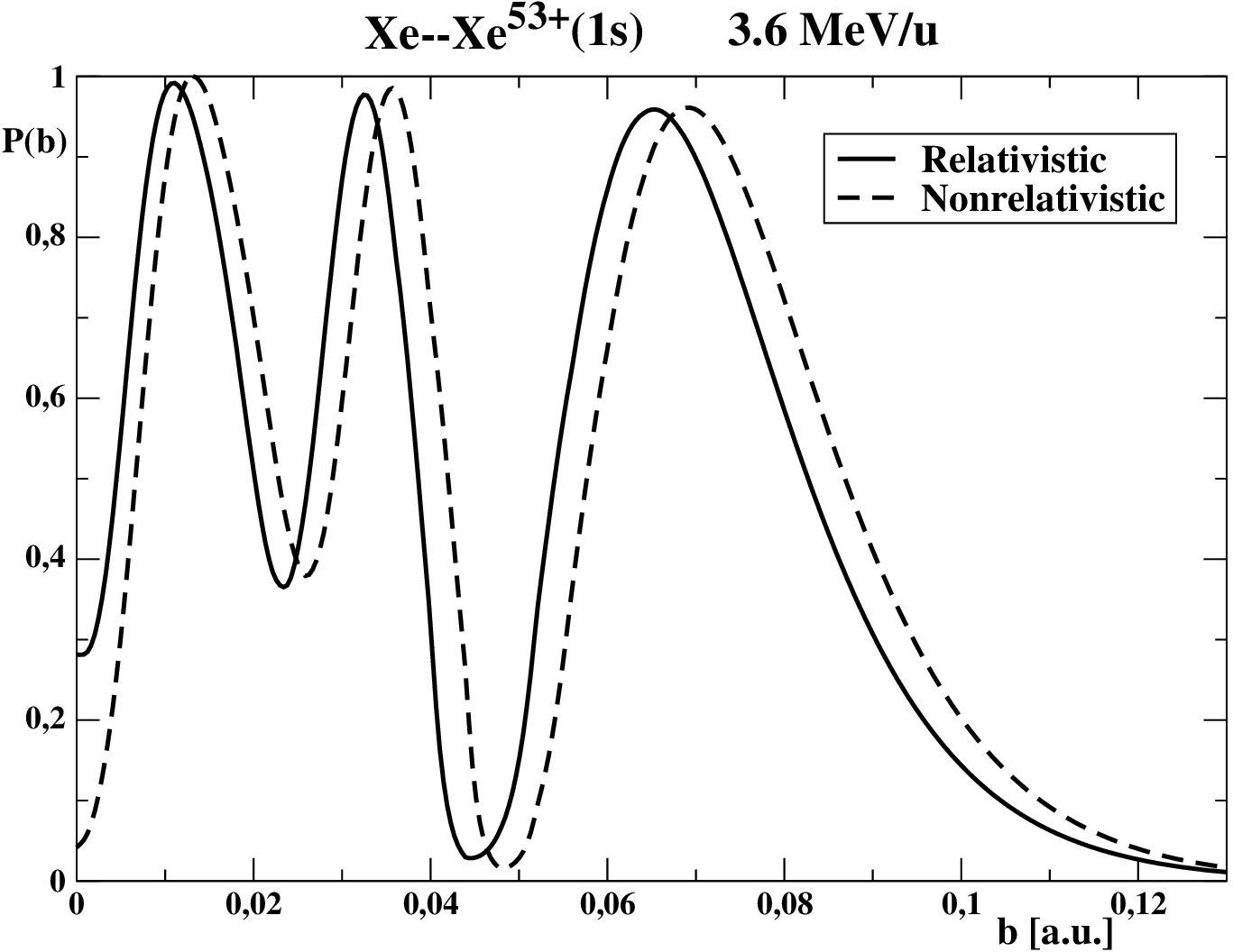}
\caption{The probability $P(b)$ of the Xe K-shell-vacancy production 
in the  Xe-Xe$^{53+}(1s)$ collision
as a
function of the impact parameter $b$.
The solid and dashed lines present relativistic and nonrelativistic results,
respectively.
}
\label{Fig:XeXe_3.6_nonrel}
\end{figure}

To investigate the role of the relativistic effects we performed
the same calculations for the  Xe-Xe$^{53+}(1s)$ collision 
in the nonrelativistic limit by multiplying the standard
value of the speed of light by the factor 1000. The obtained relativistic
and nonrelativistic results are presented in Fig.~\ref{Fig:XeXe_3.6_nonrel}.
As one can see from the figure,
the oscillatory behavior of both curves is the same but the nonrelativistic
curve is shifted toward higher impact parameters.
%
\section{Conclusion}
In this paper the method that was previously developed 
for evaluation of the electron-excitation and charge-transfer processes 
in collisions of  H-like ions with  bare nuclei has been extended 
to collisions of ions 
with neutral atoms. The extention is
based on the active electron approximation, in which the interaction of the
active electron with the passive electrons
is accounted for by the screening DFT potential.

The method developed has been applied to evaluate the K-vacancy production
and the K-K charge transfer in the low-energy Ne~--~F$^{8+}$ collision.
The results of the calculation 
are compared with available experimental data and with theoretical 
calculations by other authors.
The influence
of the relativistic effects on the K-vacancy production probability
is investigated for the Xe~--~Xe$^{53+}$ collision.
It is demonstrated that the relativistic and nonrelativistic probabilities as
functions of the impact parameter exhibit the same
oscillatory behavior at low energies but the relativistic curves
are shifted toward lower impact parameters compared to the
nonrelativistic ones. 

In our further investigation we plan to continue calculations of low-energy
heavy-ion collisions that are of interest  for current and nearest future experiments
at GSI and  FAIR facilities in Darmstadt. Special attention will be
paid to the critical regime, when the ground-state level of the united quasimolecule
dives into the negative-energy Dirac continuum.

%
\clearpage
\section{Acknowledgments}
This work was supported by DFG (Grants No. PL 254/7-1 and VO 1707/1-1), by RFBR
(Grants No. 10-02-00450 and  No. 11-02-00943-a), by GSI, by DAAD, by the Ministry of Education and Science of
Russian Federation (Grant No. P1334). The work of A.I.B. and I.A.M. was also supported by the Dynasty
foundation.
Y.S.K. and  A.I.B. acknowledge financial support by the FAIR--Russia Research 
Center.

%
\clearpage

%
\end {document}